\documentclass[twocolumn,twoside,slac_two]{revtex4}
\usepackage{graphicx}
\usepackage{fancyhdr}
\begin{document}

\title{
Development of FTK Architecture:

A Fast Hardware Track Trigger for the ATLAS Detector
}

%

\author{A. Annovi}
\affiliation{INFN Frascati}
\author{M. Beretta}
\affiliation{INFN Frascati}
\author{E. Bossini}
\affiliation{Univ. and INFN of Pisa}
\author{A. Boveia}
\affiliation{Univ. of Chicago}
\author{E. Brubaker}
\affiliation{Univ. of Chicago}
\author{F. Canelli}
\affiliation{Univ. of Chicago}
\author{V. Cavasinni}
\affiliation{Univ. and INFN of Pisa}
\author{F. Crescioli}
\affiliation{Univ. and INFN of Pisa}
\author{H. DeBerg}
\affiliation{Univ. of Illinois at Urbana-Champaign}
\author{M. Dell'Orso}
\affiliation{Univ. and INFN of Pisa}
\author{M. Dunford}
\affiliation{Univ. of Chicago}
\author{M. Franklin}
\affiliation{Harvard Univ.}
\author{P. Giannetti}
\affiliation{Univ. and INFN of Pisa}
\author{A. Kapliy*}
\affiliation{Univ. of Chicago}
\author{Y.K. Kim}
\affiliation{Univ. of Chicago}
\author{N. Kimura}
\affiliation{Waseda University}
\author{P. Laurelli}
\affiliation{INFN Frascati}
\author{A. McCarn}
\affiliation{Univ. of Illinois at Urbana-Champaign}
\author{C. Melachrinos}
\affiliation{Univ. of Chicago}
\author{C. Mills}
\affiliation{Harvard Univ.}
\author{M. Neubauer}
\affiliation{Univ. of Illinois at Urbana-Champaign}
\author{J. Proudfoot}
\affiliation{Argonne National Lab}
\author{M. Piendibene}
\affiliation{Univ. and INFN of Pisa}
\author{G. Punzi}
\affiliation{Univ. and INFN of Pisa}
\author{F. Sarri}
\affiliation{Univ. and INFN of Pisa}
\author{L. Sartori}
\affiliation{Univ. and INFN of Pisa}
\affiliation{Marie Curie Fellowship}
\author{M. Shochet}
\affiliation{Univ. of Chicago}
\author{L. Tripiccione}
\affiliation{Univ. and INFN Ferrara}
\author{J. Tuggle}
\affiliation{Univ. of Chicago}
\author{I. Vivarelli}
\affiliation{Univ. and INFN of Pisa}
\author{G. Volpi}
\affiliation{Univ. and INFN of Pisa}
\author{K. Yorita}
\affiliation{Waseda University}
\author{J. Zhang}
\affiliation{Argonne National Lab}

\begin{abstract}
As the LHC luminosity is ramped up to the design level of $10^{34}$ $cm^{-2}s^{-1}$ and beyond, the high rates, multiplicities, and energies of particles seen by the detectors will pose a unique challenge. Only a tiny fraction of the produced collisions can be stored on tape and immense real-time data reduction is needed. An effective trigger system must maintain high trigger efficiencies for the physics we are most interested in, and at the same time suppress the enormous QCD backgrounds. This requires massive computing power to minimize the online execution time of complex algorithms. A multi-level trigger is an effective solution for an otherwise impossible problem.

\parindent=10pt
The Fast Tracker (FTK) is a proposed upgrade to the ATLAS trigger system 
that will operate at full Level-1 output rates and provide high quality tracks reconstructed 
over the entire detector by the start of processing in Level-2.
FTK solves the combinatorial challenge inherent to tracking by exploiting the massive parallelism of Associative Memories (AM)
that can compare inner detector hits to millions of pre-calculated patterns simultaneously.
The tracking problem within matched patterns is further simplified by using pre-computed linearized fitting constants and leveraging
fast DSP's in modern commercial FPGA's. Overall, FTK is able to compute the helix parameters for all tracks in an event 
and apply quality cuts in approximately one millisecond. By employing a pipelined architecture, FTK is able to
continuously operate at Level-1 rates without deadtime.

\parindent=10pt
The system design is defined and studied using ATLAS full simulation. Reconstruction quality is evaluated for single muon events with zero pileup, as well as WH events at the LHC design luminosity. FTK results are compared with the tracking capability of an offline algorithm.

\end{abstract}

\maketitle

\thispagestyle{fancy}


\section{Introduction}
The Large Hadron Collider will collide proton bunches every 25 nanoseconds with a center-of-mass energy of 14 TeV. At the design luminosity, each collision on average produces 23 minimum-bias interactions that result in high detector occupancy and create a challenging environment for event readout and reconstruction. On one hand, limited data store bandwidth demands a significant online rate reduction of 5-6 orders of magnitude. On the other hand, events with interesting physics signatures must be selected very efficiently from the vast LHC background.

The ATLAS experiment employs a sophisticated three-level trigger system to achieve these goals~\cite{tdr1,tdr2}. Level-1 selection is performed in dedicated hardware that uses coarse-granularity information from calorimeters and muon spectrometers to apply cuts to a variety of objects, such as jets, muons, electromagnetic clusters, and missing energy. Although adding tracking information at this stage would be extremely beneficial to discern certain physics objects, the small timing window available to Level-1 (2.5 $\mu s$) makes this currently impossible. Moreover, even projected CPU farms that constitute the Level-2 trigger cannot perform global track reconstruction within their time budget of about 10 ms. Instead, Level-2 does limited tracking inside Regions of Interest (ROI) identified by the Level-1 trigger.

FastTracker (FTK)~\cite{alb01,alb02} is a proposed dedicated hardware track processor inspired by the Silicon Vertex Trigger (SVT)~\cite{svt} from the CDF detector. FTK operates in parallel with the normal silicon detector readout following each Level-1 trigger and reconstructs tracks over the entire detector volume (up to $|\eta|$ of 2.5) in under a millisecond. Having tracks available by the beginning of Level-2 processing allows reduced Level-1 $p_{T}$ thresholds since the Level-2 trigger is able to reject non-interesting events more quickly. Furthermore, since Level-2 is freed up from tracking, the extra processing time becomes available for more advanced online algorithms.

\section{Data Flow}
\begin{figure}[h]
\centering
\includegraphics[width=80mm]{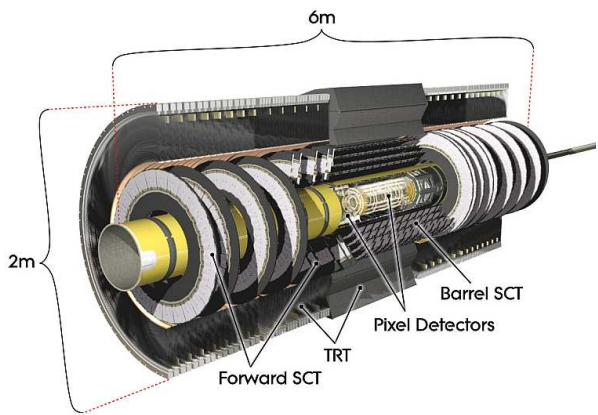}
\caption{ATLAS Inner Detector. FTK uses data from the Pixel and SCT detectors.} \label{inner_detector}
\end{figure}

\begin{figure}[h]
\centering
\includegraphics[width=80mm]{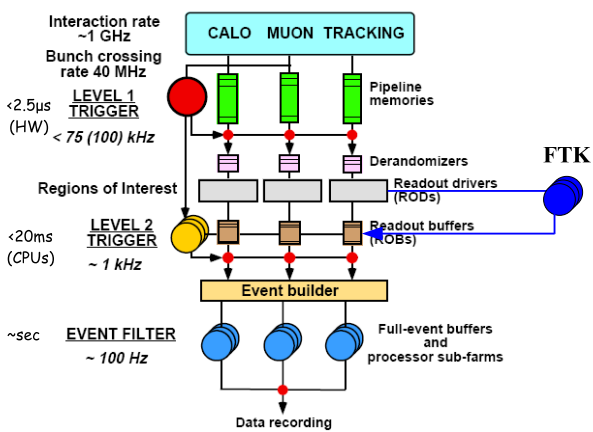}
\caption{ATLAS trigger system and its integration with FTK.} \label{trigger_system}
\end{figure}

In order to perform tracking with good efficiency and resolution, FTK uses data from the two inner-most subsystems of the ATLAS Inner Detector~\cite{indet} (Fig.~\ref{inner_detector}). Pixel sensors provide a two-dimensional measurement of hit position and include over 80 million readout channels. SCT layers are arranged in pairs of axial and narrow-angle stereo strips and consist of 6 million channels. A typical track passes through 11 detector layers and can be reconstructed from the 14 coordinate measurements ($3 \cdot 2$ from two-dimensional pixels and $8$ from SCT).
Custom-designed optical splitters duplicate Pixel and SCT readout data and send it to FTK - at full Level-1 rate. Since FTK non-invasively eavesdrops on the Inner Detector data, it easily integrates with the current ATLAS trigger system (Fig.~\ref{trigger_system}).

\begin{figure}[h]
\centering
\includegraphics[width=80mm]{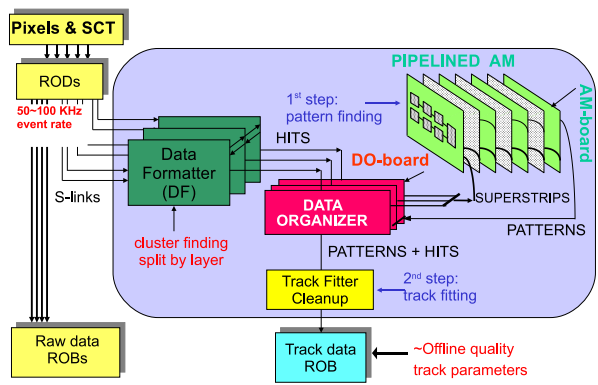}
\caption{FTK Data Flow} \label{data_flow}
\end{figure}

In order to increase the overall throughput of the system, FTK splits incoming data into 8 or more {\it regions} in $\phi$ with sufficient overlap to account for inefficiencies at the edges. Each region is served by a separate crate that consists of the subsystems shown in Fig.~\ref{data_flow}. Raw Pixel and SCT data are first received by the Data Formatter that uses fast FPGA's to perform clustering~\cite{clust}. Clustered hits proceed into the Data Organizer, where they are buffered and merged into coarse superstrips used in pattern recognition. The superstrips are then sent into a pipelined array of Associative Memory boards that perform fast pattern finding using a pre-calculated table of particle trajectories. Matched patterns are reconnected with their corresponding full-resolution hits in the Data Organizer and sent to the Track Fitters. After removal of duplicate tracks, the resulting set of FTK tracks is saved in the track data Read-Out Buffer (ROB) and made available to Level-2 processors.

\section{Pattern Recognition in Associative Memories}
\begin{figure}[h]
\centering
\includegraphics[width=80mm]{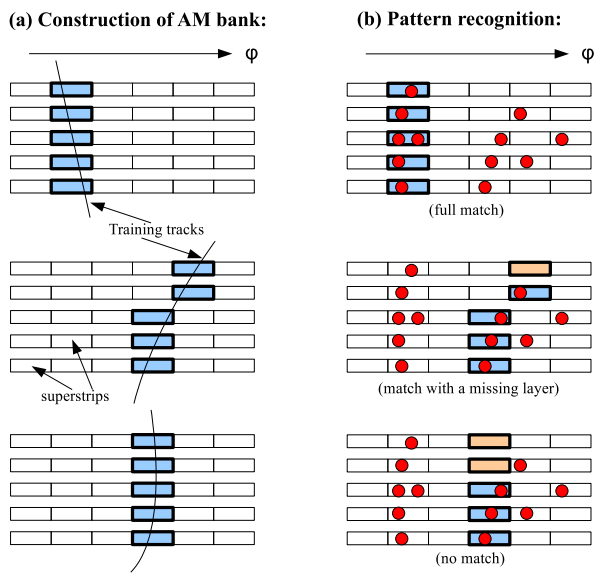}
\caption{Associative Memory and pattern bank operation} \label{am_bank}
\end{figure}

Luminosities above $10^{34} cm^{-2}s^{-1}$ combined with 86 million readout channels create a unique combinatorial challenge for tracking. FTK overcomes this with the help of specialized hardware called Associative Memory (AM) - a massive, ultra-fast lookup table that enumerates all realistic particle trajectories (patterns) through the 11 detector layers~\cite{vlsi}. In order to keep the size of the trajectory lookup under control, detector hits are merged into coarse-resolution {\it superstrips} having a width of a few millimeters\footnote{These coarse superstrips are only used in the pattern recognition stage; all final fits are performed with full resolution hits.}. The pattern bank is precalculated either from single track Monte-Carlo or from real data events (Fig.~\ref{am_bank}(a)) and stored in the AM boards.

Each pattern in the AM includes its own comparison logic. When hits from a given event enter AM boards, they are simultaneously compared with millions of pre-stored patterns. In order to account for inefficiencies in individual detector layers, FTK also matches patterns with one missing layer (Fig.~\ref{am_bank}(b)).
\begin{figure}[h]
\centering
\includegraphics[width=80mm]{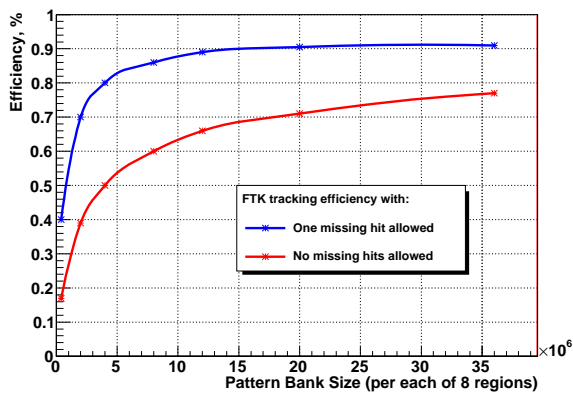}
\caption{Efficiency as a function of pattern bank size for a particular choice of superstrip widths.} \label{pattern_efficiency}
\end{figure}

Fig.~\ref{pattern_efficiency} shows the efficiency for muon track reconstruction as a function of pattern bank size for the pixel superstrip size of 3 mm and SCT superstrip size of 5 mm. With missing-layer pattern matching, efficiency quickly rises to the $90\%$ level.

\begin{figure}[h]
\centering
\includegraphics[width=80mm]{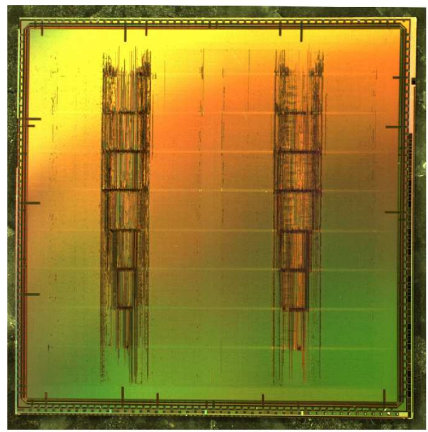}
\caption{AM chip currently used in CDF detector at Fermilab} \label{am_chip}
\end{figure}

Fig.~\ref{am_chip} shows the current AM chip used in the CDF detector at Fermilab. It uses 0.18 $\mu m$ custom cells and contains up to 2,500 patterns in the 12-layer configuration. Using standard cells with 90 nm technology, the capacity can be increased to 10,000 patterns per chip, with another factor of two gain possible with a custom cell design.

\section{Track Fitting}
FTK computes five track helix parameters (curvature, $d_{0}$ etc) and a $\chi^2$ quality of fit from the {\it full resolution} hits within each matched pattern. Since patterns are constructed from reduced-granularity superstrips, multiple full-resolution hits can belong to a given superstrip. This results in some ambiguity, which is resolved by fitting all combinations within the superstrips.

Performing full $\chi^2$ minimization with respect to five parameters is an extremely slow procedure. Instead, FTK reduces the track fitting problem to a set of scalar products, which can be computed efficiently using DSP units in modern commercial FPGA's. This is done by arranging geometrically similar patterns into a number of groups (called {\it sectors}), so that within each sector the relationship between hit positions ($x_{j}$) and track parameters ($p_{i}$) is approximately linear:
\begin{equation}
  p_{i} = \sum_{j=1}^{14} c_{ij} \cdot x_{j} + q_{i}
\label{eq-linear}
\end{equation}
The fitting coefficients for each sector are precomputed from the same training data that was used in pattern generation. An added advantage of this approach is that when real detector hits are used in training, misalignments and other detector effects are automatically taken into account.

Overall, the linearized approach allows FTK to achieve near-offline resolution with a fitting rate of about 1 fit per nanosecond.
\section{Performance}
\begin{table}[h]
\begin{center}
\caption{- FTK and offline resolutions for tracks with $p_{T}>1~GeV$ and $|\eta|<1$}
\begin{tabular}{|l|c|c|}
\hline \textbf{Track Parameter} & \textbf{$\sigma$(FTK)} & \textbf{$\sigma$(offline)} \\
\hline $1/(2p_{T}) [c/GeV]$ & $7.4 \cdot 10^{-3}$ & $6.6 \cdot 10^{-3}$ \\
\hline $\phi [rad]$ & $9.5 \cdot 10^{-4}$ & $6.3 \cdot 10^{-4}$ \\
\hline $d_{0} [cm]$ & $5.3 \cdot 10^{-3}$ & $3.3 \cdot 10^{-3}$ \\
\hline $cot(\theta)$ & $2.0 \cdot 10^{-3}$ & $1.4 \cdot 10^{-3}$ \\
\hline $z_{0} [mm]$ & $2.1 \cdot 10^{-2}$ & $1.9 \cdot 10^{-2}$ \\
\hline
\end{tabular}
\label{resolution_table}
\end{center}
\end{table}

\begin{figure}[h]
\centering
\includegraphics[width=80mm]{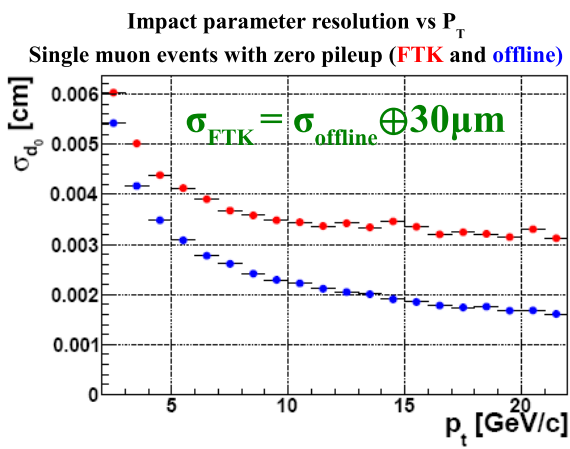}
\caption{FTK reconstruction performance for zero-pileup muons} \label{resolution_muons}
\end{figure}

\begin{figure}[h]
\centering
\includegraphics[width=80mm]{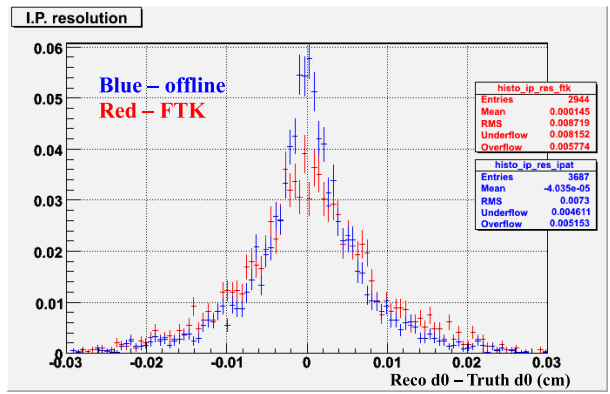}
\caption{Impact Parameter resolution for all primary tracks with $p_{T} > $ 1 GeV in WH (H $\rightarrow$ bb) events at the design luminosity - comparison between FTK and offline tracking} \label{resolution_wh}
\end{figure}

Table~\ref{resolution_table} compares FTK track parameter resolutions for muons with an offline algorithm. Overall performances are comparable; in particular, the FTK impact parameter resolution is equal to that of offline with an additional 30 microns added in quadrature (Fig.~\ref{resolution_muons}). FTK reconstruction remains robust in higher pileup environments (Fig.~\ref{resolution_wh}).

Preliminary timing estimates obtained from $10^{34}$ $cm^{-2}s^{-1}$ simulation show that FTK is able to reconstruct complex events in about 1 ms. At higher luminosities, the number of fits performed in Track Fitters can become excessively large. This can be dramatically reduced by narrowing the superstrip width or modifying the pattern recognition and fit strategy. Several potential approaches have been identified and simulated, promising to reduce FTK processing time by more than an order of magnitude.

\subsection{Physics Implications}
\begin{figure}[h]
\centering
\includegraphics[width=80mm]{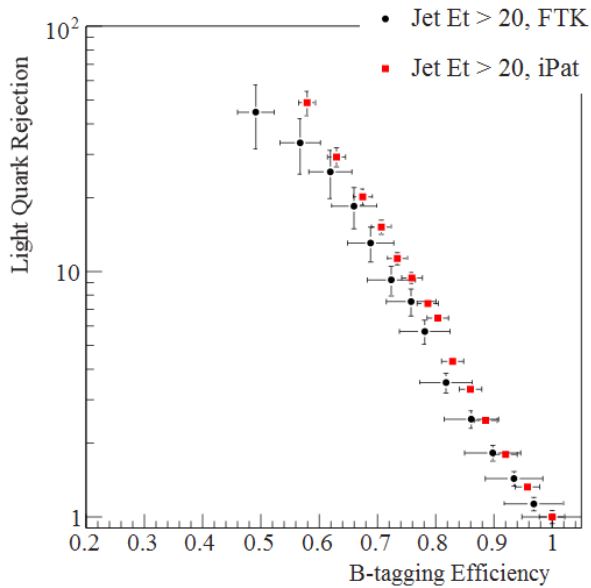}
\caption{Likelihood ratio b-tagging performance with FTK and offline tracks using the same algorithm.} \label{monica_plot}
\end{figure}

Offline-quality b-tagging efficiency and light quark rejection can be achieved by using the savings in tracking time to apply more sophisticated b-tagging algorithms at Level-2. Fig.~\ref{monica_plot} compares b-tagging performance of FTK tracks with that of offline tracks using a simple transverse impact parameter likelihood algorithm. FTK tracks provide tagging performance competitive with offline tracks.

Studies are underway to quantify the efficiency and background rejection achieved with FTK tracks with respect to other high-Pt Level-2 objects, including tau-jets and isolated leptons.

\section{Conclusions and Outlook}
FTK performs global track reconstruction at the full Level-1 trigger rate and naturally integrates with the current ATLAS data acquisition system. Using massively parallel Associative Memories, it will provide a complete list of three-dimensional tracks at the beginning of Level-2 processing, including tracks outside of the Regions of Interest. The extra time saved by FTK can be used in Level-2 to apply more advanced algorithms and ultimately extend the physics reach of the detector.

FTK robustly and quickly reconstructs tracks at the LHC design luminosity and produces efficiencies and resolutions on par with offline tracking. Studies are undergoing to evaluate the performance of the system under higher pileup conditions ($3 \cdot 10^{34}$ $cm^{-2}s^{-1}$).

We expect to produce a Technical Design Report in the fall of 2009 and the first board prototypes in 2010. The entire system will be ready in time for the LHC Phase I shutdown.

\bigskip 

\end{document}